\def\year{2022}\relax
\documentclass[letterpaper]{article} %
\usepackage{aaai22}  %
\usepackage{times}  %
\usepackage{helvet}  %
\usepackage{courier}  %
\usepackage[hyphens]{url}  %
\usepackage{graphicx} %
\usepackage{natbib}  %
\usepackage{caption} %
\DeclareCaptionStyle{ruled}{labelfont=normalfont,labelsep=colon,strut=off} %
\usepackage{comment}
\usepackage{graphicx}
\usepackage{amsmath}
\usepackage{color}
\usepackage{appendix}
\usepackage{lipsum}
\usepackage{wrapfig}
\usepackage{multirow}
\usepackage{subcaption}

\usepackage{enumitem}
\usepackage[linesnumbered,ruled]{algorithm2e}
\usepackage{mathtools,amssymb,mathrsfs}
\usepackage{booktabs}

\newcommand{\eg}{\textit{e}.\textit{g}.}

\newcommand{\defeq}{\coloneqq}
\newcommand{\grad}{\nabla}
\newcommand{\E}{\mathbb{E}}

\newcommand{\Eb}[2]{\E_{#1}\!\left[#2\right]}

\newcommand{\kl}[2]{D_{\mathrm{KL}}\!\left(#1 ~ \| ~ #2\right)}

\newcommand{\bI}{\mathbf{I}}

\newcommand{\bzero}{\mathbf{0}}

\newcommand{\by}{\mathbf{y}}
\newcommand{\bz}{\mathbf{z}}

\newcommand{\bepsilon}{{\boldsymbol{\epsilon}}}
\newcommand{\bmu}{{\boldsymbol{\mu}}}

\title{DiffSinger: Singing Voice Synthesis via Shallow Diffusion Mechanism}
\author{
    Jinglin Liu\thanks{Equal contribution.}, %
    Chengxi Li\footnotemark[1], %
    Yi Ren\footnotemark[1], %
    Feiyang Chen, %
    Zhou Zhao\thanks{Corresponding author} %
}
\affiliations{
    Zhejiang University \\
    \{jinglinliu,chengxili,rayeren,zhaozhou\}@zju.edu.cn, 
    chenfeiyangai@gmail.com

}

\usepackage{bibentry}

\begin{document}

\maketitle

\begin{abstract}
Singing voice synthesis (SVS) systems are built to synthesize high-quality and expressive singing voice, in which the acoustic model generates the acoustic features (\eg, mel-spectrogram) given a music score. Previous singing acoustic models adopt a simple loss (\eg, L1 and L2) or generative adversarial network (GAN) to reconstruct the acoustic features, while they suffer from over-smoothing and unstable training issues respectively, which hinder the naturalness of synthesized singing. 
In this work, we propose DiffSinger, an acoustic model for SVS based on the diffusion probabilistic model. DiffSinger is a parameterized Markov chain that iteratively converts the noise into mel-spectrogram conditioned on the music score. By implicitly optimizing variational bound, DiffSinger can be stably trained and generate realistic outputs. 
To further improve the voice quality and speed up inference, we introduce a shallow diffusion mechanism to make better use of the prior knowledge learned by the simple loss. Specifically, DiffSinger starts generation at a shallow step smaller than the total number of diffusion steps, according to the intersection of the diffusion trajectories of the ground-truth mel-spectrogram and the one predicted by a simple mel-spectrogram decoder. Besides, we propose boundary prediction methods to locate the intersection and determine the shallow step adaptively.
The evaluations conducted on a Chinese singing dataset demonstrate that DiffSinger outperforms state-of-the-art SVS work. Extensional experiments also prove the generalization of our methods on text-to-speech task (DiffSpeech). Audio samples: \url{https://diffsinger.github.io}. Codes: \url{https://github.com/MoonInTheRiver/DiffSinger}.
\end{abstract}

\section{Introduction} \label{sec:intro}
Singing voice synthesis (SVS) which aims to synthesize natural and expressive singing voice from musical score~\citep{wu2020adversarially}, increasingly draws attention from the research community and entertainment industries~\citep{zhang2020durian}. The pipeline of SVS usually consists of an acoustic model to generate the acoustic features (\eg, mel-spectrogram) conditioned on a music score, and a vocoder to convert the acoustic features to waveform~\citep{nakamura2019singing,lee2019adversarially,blaauw2020sequence,ren2020deepsinger,chen2020hifisinger}\footnote{A music score consists of lyrics, pitch and duration.}.

Previous singing acoustic models mainly utilize simple loss (\eg, L1 or L2) to reconstruct the acoustic features. However, this optimization is based on the incorrect uni-modal distribution assumptions, leading to blurry and over-smoothing outputs. Although existing methods endeavor to solve this problem by generative adversarial network (GAN)~\cite{lee2019adversarially,chen2020hifisinger}, training an effective GAN may occasionally fail due to the unstable discriminator. These issues hinder the naturalness of synthesized singing.

Recently, a highly flexible and tractable generative model, diffusion probabilistic model (a.k.a. diffusion model)~\cite{sohl2015deep,Ho2020ddpm,song2021denoising} emerges. Diffusion model consists of two processes: diffusion process and reverse process (also called denoising process). The diffusion process is a Markov chain with fixed parameters (when using the certain parameterization in \cite{Ho2020ddpm}), which converts the complicated data into isotropic Gaussian distribution by adding the Gaussian noise gradually; while the reverse process is a Markov chain implemented by a neural network, which learns to restore the origin data from Gaussian white noise iteratively. Diffusion model can be stably trained by implicitly optimizing variational lower bound (ELBO) on the data likelihood. It has been demonstrated that diffusion model can produce promising results in image generation~\cite{Ho2020ddpm,song2021denoising} and 
neural vocoder~\cite{chen2021wavegrad,kong2021diffwave} fields.

In this work, we propose DiffSinger, an acoustic model for SVS based on diffusion model, which converts the noise into mel-spectrogram conditioned on the music score. DiffSinger can be efficiently trained by optimizing ELBO, without adversarial feedback, and generates realistic mel-spectrograms strongly matching the ground truth distribution. 

To further improve the voice quality and speed up inference, we introduce a shallow diffusion mechanism to make better use of the prior knowledge learned by the simple loss. Specifically, we find that there is an intersection of the diffusion trajectories of the ground-truth mel-spectrogram $M$ and the one predicted by a simple mel-spectrogram decoder $\widetilde{M}$\footnote{Here we use a traditional acoustic model based on feed-forward Transformer~\cite{ren2021fastspeech,blaauw2020sequence}, which is trained by L1 loss to reconstruct mel-spectrogram.}: sending $M$ and $\widetilde{M}$ into the diffusion process could result in similar distorted mel-spectrograms, when the diffusion step is big enough (but not reaches the deep step where the distorted mel-spectrograms become Gaussian white noise). Thus, in the inference stage we 1) leverage the simple mel-spectrogram decoder to generate $\widetilde{M}$; 2) calculate the sample at a shallow step $k$ through the diffusion process: $\widetilde{M}_k$\footnote{$K < T$, where T is the total number of diffusion steps. $\widetilde{M}_k$ can be calculated in closed form time~\cite{Ho2020ddpm}.}; and 3) start reverse process from $\widetilde{M}_k$ rather than Gaussian white noise, and complete the process by $k$ iteration denoising steps~\cite{score2011vincent,song2019generative,Ho2020ddpm}. Besides, we train a boundary prediction network to locate this intersection and determine the $k$ adaptively. The shallow diffusion mechanism provides a better start point than Gaussian white noise and alleviates the burden of the reverse process, which improves the quality of synthesized audio and accelerates inference. %

Finally, since the pipeline of SVS resembles that of text-to-speech (TTS) task, we also build DiffSpeech adjusting from DiffSinger for generalization. The evaluations conducted on a Chinese singing dataset demonstrate the superiority of DiffSinger (0.11 MOS gains compared with a state-of-the-art acoustic model for SVS~\cite{wu2020adversarially}), and the effectiveness of our novel mechanism (0.14 MOS gains, 0.5 CMOS gains and 45.1\% speedup with shallow diffusion mechanism). The extensional experiments of DiffSpeech on TTS task prove the generalization of our methods (0.24/0.23 MOS gains compared with FastSpeech 2~\cite{ren2021fastspeech} and Glow-TTS~\cite{kim2020glow} respectively).
The contributions of this work can be summarized as follows:
\begin{itemize}%
\item We propose DiffSinger, which is the first acoustic model for SVS based on diffusion probabilistic model. DiffSinger addresses the over-smoothing and unstable training issues in previous works.
\item We propose a shallow diffusion mechanism to further improve the voice quality, and accelerate the inference.
\item The extensional experiments on TTS task (DiffSpeech) prove the generalization of our methods.
\end{itemize}

\section{Diffusion Model} \label{sec:ddpm}
\begin{figure}[!h]
	\centering
	\includegraphics[width=0.49\textwidth]{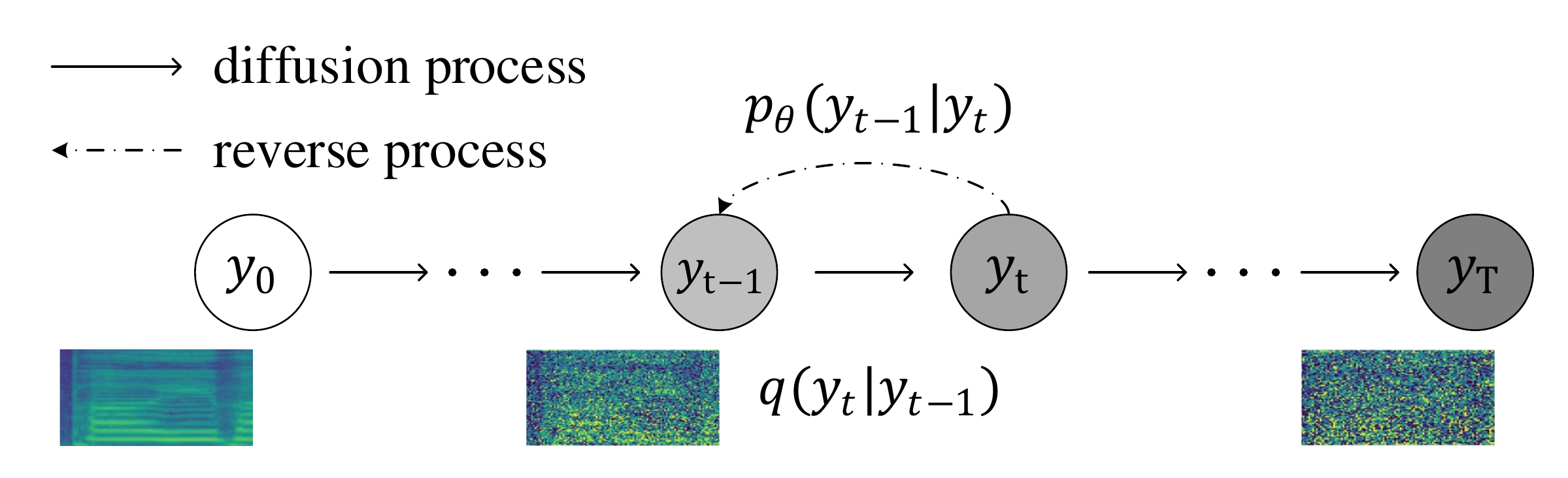}
	\caption{The directed graph for diffusion model.}
	\label{fig:two_process}
\end{figure}

In this section, we introduce the theory of diffusion probabilistic model~\cite{sohl2015deep,Ho2020ddpm}. The full proof can be found in previous works \cite{Ho2020ddpm,kong2021diffwave,song2021denoising}. A diffusion probabilistic model converts the raw data into Gaussian distribution gradually by a diffusion process, and then learns the reverse process to restore the data from Gaussian white noise~\cite{sohl2015deep}. These processes are shown in Figure~\ref{fig:two_process}. 
\paragraph{Diffusion Process}
Define the data distribution as $q(\by_0)$, and sample $\by_0\sim q(\by_0)$. The diffusion process is a Markov chain with fixed parameters~\cite{Ho2020ddpm}, which converts $\by_0$ into the latent $\by_T$ in $T$ steps:
\begin{equation*}
q(\by_{1:T} | \by_0) \defeq \prod_{t=1}^T q(\by_T | \by_{t-1} ).
\end{equation*}
At each diffusion step $t \in [1, T]$, a tiny Gaussian noise is added to $\by_{t-1}$ to obtain $\by_t$, according to a variance schedule $\beta = \{\beta_1, \dotsc, \beta_T\}$:
\begin{equation*}
q(\by_t|\by_{t-1}) \defeq \mathcal{N}(\by_t;\sqrt{1-\beta_t}\by_{t-1},\beta_t \bI).
\end{equation*}
If $\beta$ is well designed and $T$ is sufficiently large, then $q(y_T)$ is nearly an isotropic Gaussian distribution~\cite{Ho2020ddpm,nichol2021improved}. Besides, there is a special property of diffusion process that $q(\by_t|\by_0)$ can be calculated in closed form in $O(1)$ time~\cite{Ho2020ddpm}:
\begin{align}
	q(\by_t|\by_0) = \mathcal{N}(\by_t; \sqrt{\bar\alpha_t}\by_0, (1-\bar\alpha_t)\bI), \label{eq:one_step_noise}
\end{align}
where $\bar\alpha_t \defeq \prod_{s=1}^t \alpha_s$, $\alpha_t \defeq 1-\beta_t$.
\paragraph{Reverse Process}
The reverse process is a Markov chain with learnable parameters $\theta$ from $\by_T$ to $\by_0$. Since the exact reverse transition distribution $q(\by_{t-1}| \by_t)$ is intractable, we approximate it by a neural network with parameters $\theta$ ($\theta$ is shared at every $t$-th step):
\begin{equation} \label{eq:reverse_step}
p_\theta(\by_{t-1}|\by_t) \defeq \mathcal{N}(\by_{t-1}; \bmu_\theta(\by_t, t), \sigma_t^2\bI). %
\end{equation}
Thus the whole reverse process can be defined as:
\begin{equation*}
p_\theta(\by_{0:T}) \defeq p(\by_T)\prod_{t=1}^T p_\theta(\by_{t-1}|\by_t).
\end{equation*}

\begin{figure*}[!ht]
    \centering
    \begin{subfigure}{0.41\textwidth}
    	\centering
    	\includegraphics[width=\textwidth,clip=true]{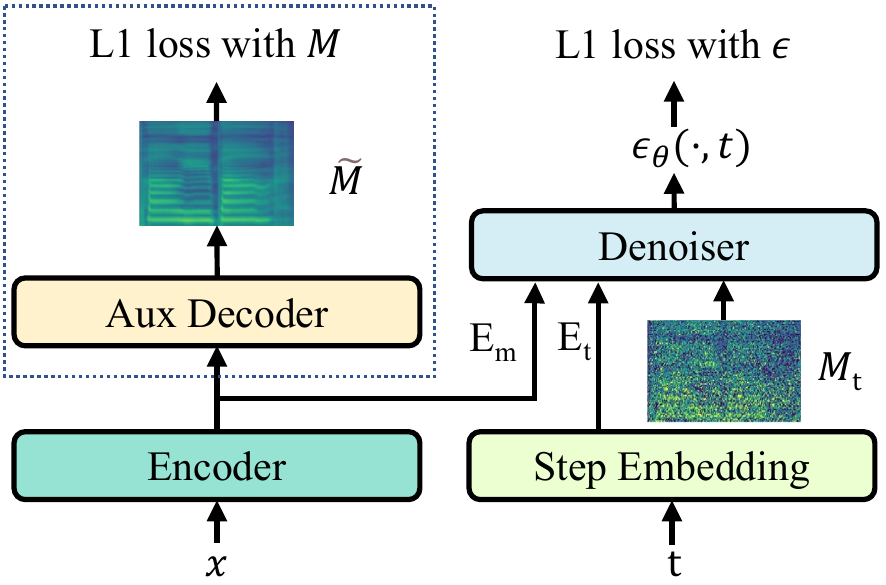}
    	\caption{The training procedure of DiffSinger.}
    	\label{fig:main_fig_sub1}
    \end{subfigure}
    \hspace{0.5cm}
    \begin{subfigure}{0.43\textwidth}
    	\centering
    	\includegraphics[width=\textwidth,clip=true]{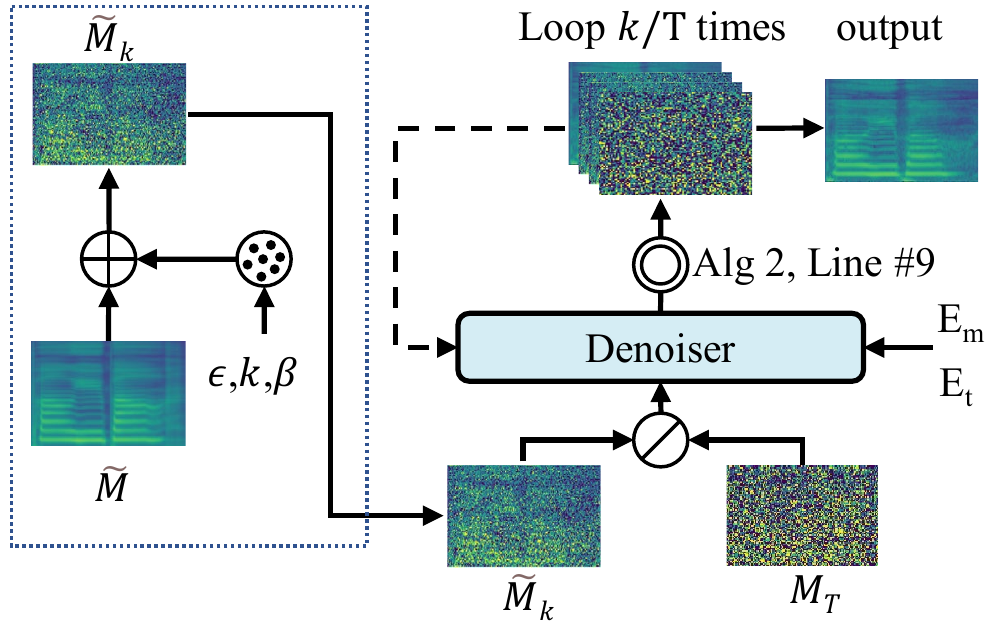}
    	\caption{The inference procedure of DiffSinger.}
    	\label{fig:main_fig_sub2}
    \end{subfigure}
    \caption{The overview of DiffSinger (with shallow diffusion mechanism in the dotted line boxes). In subfigure (a), $x$ is the music score; $t$ is the step number; $M$ means the ground truth mel-spectrogram; $\widetilde{M}$ means the blurry mel-spectrogram generated by the auxiliary decoder trained with L1 loss; $M_t$ is $M$ at the $t$-th step in the diffusion process. In subfigure (b), $M_T$ means the $M$ at $T$-th diffusion step (Gaussian white noise); $k$ is the predicted intersection boundary; there is a switch to select $M_T$ (naive version) or $\widetilde{M}_k$ (with shallow diffusion) as the start point of the inference procedure.}
    \label{fig:main_fig}
\end{figure*}

\paragraph{Training}
To learn the parameters $\theta$, we minimize a variational bound of the negative log likelihood:
\begin{equation*}
\begin{split}
& \mathbb{E}_{q(\by_0)}[-\log p_\theta(\by_0)] \geq \\ 
& \mathbb{E}_{q(\by_0, \by_1, \ldots, \by_T)}\left[\log q(\by_{1:T} | \by_0) - \log p_\theta(\by_{0:T})\right] \eqqcolon \mathbb{L}.
\end{split}
\end{equation*}
Efficient training is optimizing a random term of $\mathbb{L}$ with stochastic gradient descent~\cite{Ho2020ddpm}:
\begin{align}
\mathbb{L}_{t-1} = \kl{q(\by_{t-1}|\by_t,\by_0)}{p_\theta(\by_{t-1}|\by_t)}, \label{eq:opt_tgtA}
\end{align}
where
\begin{align*}
&q(\by_{t-1}|\by_t,\by_0) = \mathcal{N}(\by_{t-1}; \tilde\bmu_t(\by_t, \by_0), \tilde\beta_t \bI)\\
&\tilde\bmu_t(\by_t, \by_0) \defeq \frac{\sqrt{\bar\alpha_{t-1}}\beta_t }{1-\bar\alpha_t}\by_0 + \frac{\sqrt{\alpha_t}(1- \bar\alpha_{t-1})}{1-\bar\alpha_t} \by_t,
\end{align*}
where $\tilde\beta_t \defeq \frac{1-\bar\alpha_{t-1}}{1-\bar\alpha_t}\beta_t$. Eq. \eqref{eq:opt_tgtA} is equivalent to:
\begin{align}
  \mathbb{L}_{t-1} - \mathcal{C}
   = \Eb{q}{ \frac{1}{2\sigma_t^2} \|\tilde\bmu_t(\by_t,\by_0) - \bmu_\theta(\by_t, t)\|^2 }, \label{eq:opt_tgtB}
\end{align}
where $\mathcal{C}$ is a constant. And by reparameterizing Eq. \eqref{eq:one_step_noise} as $\by_t(\by_0, \bepsilon) = \sqrt{\bar\alpha_t}\by_0 + \sqrt{1-\bar\alpha_t}\bepsilon$, and choosing the parameterization:
\begin{equation} \label{eq:mu_parameterization}
    \bmu_\theta(\by_t, t) = \frac{1}{\sqrt{\alpha_t}}\left( \by_t - \frac{\beta_t}{\sqrt{1-\bar\alpha_t}} \bepsilon_\theta(\by_t, t) \right),
\end{equation} 
Eq. \eqref{eq:opt_tgtB} can be simplified to:
\begin{align}
    \Eb{\by_0, \bepsilon}{ \frac{\beta_t^2}{2\sigma_t^2 \alpha_t (1-\bar\alpha_t)}  \left\| \bepsilon - \bepsilon_\theta(\sqrt{\bar\alpha_t} \by_0 + \sqrt{1-\bar\alpha_t}\bepsilon, t) \right\|^2}. \label{eq:opt_tgtC}
\end{align}
Finally we set $\sigma_t^2$ to $\tilde\beta_t$, sample $\bepsilon\sim\mathcal{N}(\bzero,\bI)$ and $\bepsilon_\theta(\cdot)$ is the outputs of the neural network.

\paragraph{Sampling}
Sample $\by_T$ from $p(\by_T) \sim \mathcal{N}(\bzero, \bI)$ and run the reverse process to obtain a data sample.

\section{DiffSinger}
As illustrated in Figure~\ref{fig:main_fig}, DiffSinger is built on the diffusion model. Since SVS task models the conditional distribution $p_\theta(M_{0}|x)$, where $M$ is the mel-spectrogram and $x$ is the music score corresponding to $M$, we add $x$ to the diffusion denoiser as the condition in the reverse process. In this section, we first describe a naive version of DiffSinger (Section \ref{sec:naive_diffsinger}); then we introduce a novel shallow diffusion mechanism to improve the model performance and efficiency (Section \ref{sec:shallow_mechanism}); finally, we describe the boundary prediction network which can adaptively find the intersection boundary required in shallow diffusion mechanism (Section \ref{sec:boundary_prediction}).

\subsection{Naive Version of DiffSinger} \label{sec:naive_diffsinger}
In the naive version of DiffSinger (without dotted line boxes in Figure~\ref{fig:main_fig}): In the training procedure (shown in Figure~\ref{fig:main_fig_sub1}), DiffSinger takes in the mel-spectrogram at $t$-th step $M_t$ in the diffusion process and predicts the random noise $\bepsilon_\theta(\cdot)$ in Eq. \eqref{eq:opt_tgtC}, conditioned on $t$ and the music score $x$. The inference procedure (shown in Figure~\ref{fig:main_fig_sub2})  starts at the Gaussian white noise sampled from $\mathcal{N}(\bzero, \bI)$, as the previous diffusion models do~\cite{Ho2020ddpm,kong2021diffwave}. Then the procedure iterates for $T$ times to repeatedly denoise the intermediate samples with two steps: 1) predict the $\bepsilon_\theta(\cdot)$ using the denoiser; 2) obtain $M_{t-1}$ from $M_t$ using the predicted $\bepsilon_\theta(\cdot)$, according to Eq. \eqref{eq:reverse_step} and Eq. \eqref{eq:mu_parameterization}:
\begin{equation*}
    M_{t-1} = \frac{1}{\sqrt{\alpha_t}}\left( M_t - \frac{1-\alpha_t}{\sqrt{1-\bar\alpha_t}} \bepsilon_\theta(M_t, x, t) \right) + \sigma_t \bz,
\end{equation*}
where $z\sim\mathcal{N}(\bzero, \bI)$ when $t>1$, and $z=0$ when $t=1$. Finally, a mel-spectrogram $\mathcal{M}$ corresponding to $x$ could be generated.

\subsection{Shallow Diffusion Mechanism} \label{sec:shallow_mechanism}
\begin{figure}
    \centering
    
    \includegraphics[width=0.46\textwidth,]{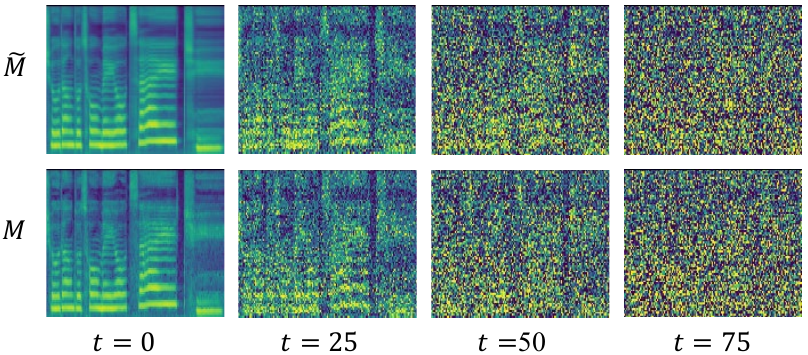}
    \caption{The mel-spectrograms at different steps in the diffusion process. The first line shows the diffusion process of mel-spectorgrams $\widetilde{M}$ generated by a simple decoder trained with L1 loss; the second line shows that of ground truth mel-spectrograms.}
    \label{fig:fake_real_mels_diffusion}
    
\end{figure}
\begin{figure}[t]
    
    \centering
    \includegraphics[width=0.46\textwidth,]{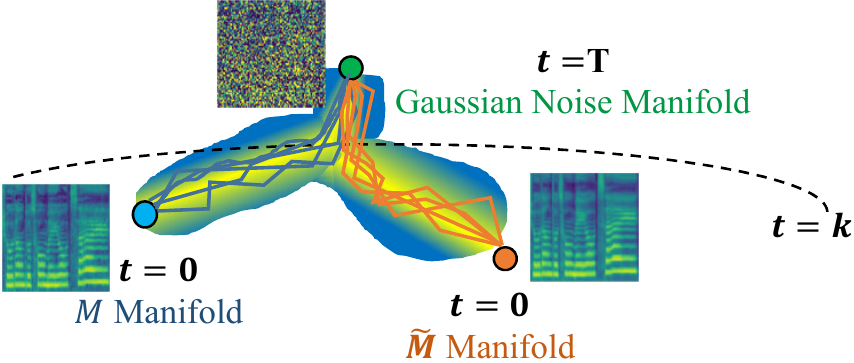} %
    \caption{The diffusion trajectories of $M$ and $\widetilde{M}$. Two distributions $q(M_t| M_0)$ and $q(\widetilde{M}_t| \widetilde{M}_0)$ become closer as $t$ increases.}
    \label{fig:manifolds}

\end{figure}

Although the previous acoustic model trained by the simple loss has intractable drawbacks, it still generates samples showing strong connection\footnote{The samples fail to maintain the variable aperiodic parameters, but they usually have a clear ``skeleton'' (harmonics) matching the ground truth.} to the ground-truth data distribution, which could provide plenty of prior knowledge to DiffSinger. To explore this connection and find a way to make better use of the prior knowledge, we conduct the empirical observation leveraging the diffusion process (shown in Figure~\ref{fig:fake_real_mels_diffusion}):
1) when $t = 0$, $M$ has rich details between the neighboring harmonics, which can influence the naturalness of the synthesized singing voice, but $\widetilde{M}$ is over-smoothing as we introduced in Section~\ref{sec:intro}; 2) as $t$ increases, samples of two process become indistinguishable. We illustrate this observation in Figure~\ref{fig:manifolds}: the trajectory from $\widetilde{M}$ manifold to Gaussian noise manifold and the trajectory from $M$ to Gaussian noise manifold intersect when the diffusion step is big enough.

Inspired by this observation, we propose the shallow diffusion mechanism: instead of starting with the Gaussian white noise, the reverse process starts at the intersection of two trajectories shown in Figure~\ref{fig:manifolds}. Thus the burden of the reverse process could be distinctly alleviated\footnote{Converting $M_k$ into $M_0$ is easier than converting $M_T$ (Gaussion white noise) into $M_0$ ($k < T$). Thus the former could improve the quality of synthesized audio and accelerates inference.}. Specifically, in the inference stage we 1) leverage an auxiliary decoder to generate $\widetilde{M}$, which is trained with L1 conditioned on the music score encoder outputs, as shown in the dotted line box in Figure~\ref{fig:main_fig_sub1}; 2) generate the intermediate sample at a shallow step $k$ through the diffusion process, as shown in the dotted line box in Figure~\ref{fig:main_fig_sub2} according to Eq. \eqref{eq:one_step_noise}: 
\begin{equation*}
\widetilde{M}_k(\widetilde{M}, \bepsilon) = \sqrt{\bar\alpha_k}\widetilde{M} + \sqrt{1-\bar\alpha_k}\bepsilon,
\end{equation*}
where $\bepsilon\sim\mathcal{N}(\bzero, \bI)$, $\bar\alpha_k \defeq \prod_{s=1}^k \alpha_s$, $\alpha_k \defeq 1-\beta_k$. If the intersection boundary $k$ is properly chosen, it can be considered that $\widetilde{M}_k$ and $M_k$ come from the same distribution; 3) start reverse process from $\widetilde{M}_k$, and complete the process by $k$ iterations denoising.
The training and inference procedures with shallow diffusion mechanism are described in Algorithm~\ref{alg:training} and \ref{alg:sampling} respectively. The theoretical proof of the intersection of two trajectories can be found in the supplement.%

\subsection{Boundary Prediction}   \label{sec:boundary_prediction}
	
We propose a boundary predictor (BP) to locate the intersection in Figure~\ref{fig:manifolds} and determine $k$ adaptively. Concretely, BP consists of a classifier and a module for adding noise to mel-spectrograms according to Eq. \eqref{eq:one_step_noise}. Given the step number $t\in[0, T]$, we label the $M_t$ as 1 and $\widetilde{M}_t$ as 0, and use cross-entropy loss to train the boundary predictor to judge whether the input mel-spectrogram at $t$ step in diffusion process comes from $M$ or $\widetilde{M}$. The training loss $\mathbb{L}_{BP}$ can be written as:
\begin{align*}
    \mathbb{L}_{BP} = - \mathbb{E}_{M\in\mathcal{Y}, t \in [0, T]} [\log BP(M_t, t) + \\
    \log (1-BP(\widetilde{M}_t, t))],
\end{align*}
where $\mathcal{Y}$ is the training set of mel-spectrograms. When BP have been trained, we determine $k$ using the predicted value of BP, which indicates the probability of a sample classified to be 1. For all $ M \in \mathcal{Y}$, we find the earliest step $k'$ where the 95\% steps $t$ in $[k', T]$ satisfies: the margin between BP($M_t$, $t$) and BP($\widetilde{M}_t$, $t$) is under the threshold. Then we choose the average of $k'$ as the intersection boundary $k$.

\begin{algorithm}[!h]
\caption{Training procedure of DiffSinger.}
\label{alg:training}
  \SetKwInOut{Input}{Input} %
    \Input{The denoiser $\bepsilon_\theta$; the intersection boundary $k$; the training set $(\mathcal{X}, \mathcal{Y})$.}
  \SetKwRepeat{Do}{do}{while}
    \Repeat{convergence}{
        Sample $(x, M)$ from $(\mathcal{X}, \mathcal{Y})$\;
        $\bepsilon\sim\mathcal{N}(\bzero,\bI)$\;
        $t \sim \mathrm{Uniform}(\{1, \dotsc, k\})$\;
        Take gradient descent step on \\
        $\quad \grad_\theta \left\| \bepsilon - \bepsilon_\theta(\sqrt{\bar\alpha_t} M + \sqrt{1-\bar\alpha_t}\bepsilon, x, t) \right\|^2$
     }
\end{algorithm}

\begin{algorithm}[!h]
\caption{Inference procedure of DiffSinger.}
\label{alg:sampling}
  \SetKwInOut{Input}{Input} %
    \Input{The denoiser $\bepsilon_\theta$; the auxiliary decoder; the intersection boundary $k$; the source testing set $\mathcal{X}$.}
    Sample $x$ from $\mathcal{X}$ as condition\;
    Generate $\widetilde{M}$ by the auxiliary decoder\;
    $\bepsilon\sim\mathcal{N}(\bzero,\bI)$\;
    $\widetilde{M}_k(\widetilde{M}, \bepsilon) = \sqrt{\bar\alpha_k}\widetilde{M} + \sqrt{1-\bar\alpha_k}\bepsilon$\;
    $M_k =  \widetilde{M}_k$\;
    \For{$t=k, k-1, ..., 1$}{
        \lIf{$t = 1$}{
        $\bz = \bzero$
        }\lElse{Sample $\bz \sim \mathcal{N}(\bzero, \bI)$}
        $M_{t-1} = \frac{1}{\sqrt{\alpha_t}}\left( M_t - \frac{1-\alpha_t}{\sqrt{1-\bar\alpha_t}} \bepsilon_\theta(M_t, x, t) \right) + \sigma_t \bz$
     }
    
\end{algorithm}

We also propose an easier trick for boundary prediction in the supplement by comparing the KL-divergence. Note that the boundary prediction can be considered as a step of dataset preprocessing to choose the hyperparameter $k$ for the whole dataset. $k$ actually can be chosen manually by brute-force searching on validation set.

\subsection{Model Structures} 
\paragraph{Encoder}
The encoder encodes the music score into the condition sequence, which consists of 1) a lyrics encoder to map the phoneme ID into embedding sequence, and a series of Transformer blocks~\cite{vaswani2017attention} to convert this sequence into linguistic hidden sequence; 2) a length regulator to expand the linguistic hidden sequence to the length of mel-spectrograms according to the duration information; and 3) a pitch encoder to map the pitch ID into pitch embedding sequence. Finally, the encoder adds linguistic sequence and pitch sequence together as the music condition sequence $E_m$ following \cite{ren2020deepsinger}. 
\paragraph{Step Embedding} 
The diffusion step $t$ is another conditional input for denoiser $\bepsilon_\theta$, as shown in Eq. \eqref{eq:opt_tgtC}. To convert the discrete step $t$ to continuous hidden, we use the sinusoidal position embedding~\cite{vaswani2017attention} followed by two linear layers to obtain step embedding $E_t$ with $C$ channels.

\paragraph{Auxiliary Decoder}
We introduce a simple mel-spectrogram decoder called the auxiliary decoder, which is composed of stacked feed-forward Transformer (FFT) blocks and generates $\widetilde{M}$ as the final outputs, the same as the mel-spectrogram decoder in FastSpeech 2~\cite{ren2021fastspeech}.
\paragraph{Denoiser}
Denoiser $\bepsilon_\theta$ takes in $M_t$ as input to predict $\bepsilon$ added in diffusion process conditioned on the step embedding $E_t$ and music condition sequence $E_m$. Since diffusion model imposes no architectural constraints~\cite{sohl2015deep,kong2021diffwave}, the design of denoiser has multiple choices. We adopt a non-causal WaveNet~\cite{vanwavenet} architecture proposed by \cite{rethage2018wavenet,kong2021diffwave} as our denoiser. The denoiser is composed of a $1\times1$ convolution layer to project $M_t$ with $H_m$ channels to the input hidden sequence $\mathcal{H}$ with $C$ channels and $N$ convolution blocks with residual connections. Each convolution block consists of 1) an element-wise adding operation which adds $E_t$ to $\mathcal{H}$; 2) a non-causal convolution network which converts $\mathcal{H}$ from $C$ to $2C$ channels; 3) a $1\times1$ convolution layer which converts the $E_m$ to $2C$ channels; 4) a gate unit to merge the information of input and conditions; and 5) a residual block to split the merged hidden into two branches with $C$ channels (the residual as the following $\mathcal{H}$ and the ``skip hidden'' to be collected as the final results), which enables the denoiser to incorporate features at several hierarchical levels for final prediction.

\paragraph{Boundary Predictor}
The classifier in the boundary predictor is composed of 1) a step embedding to provide $E_t$; 2) a ResNet~\cite{he2016deep} with stacked convolutional layers and a linear layer, which takes in the mel-spectrograms at $t$-th step and $E_t$ to classify $M_t$ and $\widetilde{M}_t$.

More details of model structure and configurations are shown in the supplement. %

\section{Experiments}
In this section, we first describe the experimental setup, and then provide the main results on SVS with analysis. Finally, we conduct the extensional experiments on TTS.
\subsection{Experimental Setup} \label{sec:setup}
\paragraph{Dataset}
Since there is no publicly available high-quality unaccompanied singing dataset, we collect and annotate a Chinese Mandarin pop songs dataset: PopCS, to evaluate our methods. PopCS contains 117 Chinese pop songs (total $\sim$5.89 hours with lyrics) collected from a qualified female vocalist. All the audio files are recorded in a recording studio. Every song is sampled at 24kHz with 16-bit quantization. To obtain more accurate music scores corresponding to the songs~\cite{lee2019adversarially}, we 1) split each whole song into sentence pieces following DeepSinger~\cite{ren2020deepsinger} and train a Montreal Forced Aligner tool (MFA)~\cite{mcauliffe2017montreal} model on those sentence-level pairs to obtain the phoneme-level alignments between song piece and its corresponding lyrics; 2) extract $F_0$ (fundamental frequency) as pitch information from the raw waveform using Parselmouth, following \cite{wu2020adversarially,blaauw2020sequence,ren2020deepsinger}. We randomly choose 2 songs for validation and testing. To release a high-quality dataset, after the paper is accepted, we clean and re-segment these songs, resulting in 1,651 song pieces, which mostly last 10$\sim$13 seconds. The codes accompanied with the access to PopCS are in \url{https://github.com/MoonInTheRiver/DiffSinger}. In this repository, we also add the extra codes out of interest, for MIDI-to-Mel, including the MIDI-to-Mel without F0 prediction/condition.

\paragraph{Implementation Details}
We convert Chinese lyrics into phonemes by pypinyin following \cite{ren2020deepsinger}; and extract the mel-spectrogram~\cite{shen2018natural} from the raw waveform; and set the hop size and frame size to 128 and 512 in respect of the sample rate 24kHz. The size of phoneme vocabulary is 61. The number of mel bins $H_m$ is 80. The mel-spectrograms are linearly scaled to the range [-1, 1], and F0 is normalized to have zero mean and unit variance. In the lyrics encoder, the dimension of phoneme embeddings is 256 and the Transformer blocks have the same setting as that in FastSpeech 2~\cite{ren2021fastspeech}. In the pitch encoder, the size of the lookup table and encoded pitch embedding are set to 300 and 256. The channel size $C$ mentioned before is set to 256. In the denoiser, the number of convolution layers N is 20 with the kernel size 3, and we set the dilation to 1 (without dilation) at each layer\footnote{You can consider setting a bigger dilation number to increase the receptive field of the denoiser. See our Github repository.}. We set $T$ to $100$ and $\beta$ to constants increasing linearly from $\beta_1 = 10^{-4}$ to $\beta_T = 0.06$. The auxiliary decoder has the same setting as the mel-spectrogram decoder in FastSpeech 2. In the boundary predictor, the number of convolutional layers is 5, and the threshold is set to 0.4 empirically.

\begin{figure*}[!ht]
    \centering
    \begin{subfigure}{0.24\textwidth}
    	\centering
    	\includegraphics[width=\textwidth,trim={1.6cm 0.5cm 1.3cm 0.5cm},clip=true]{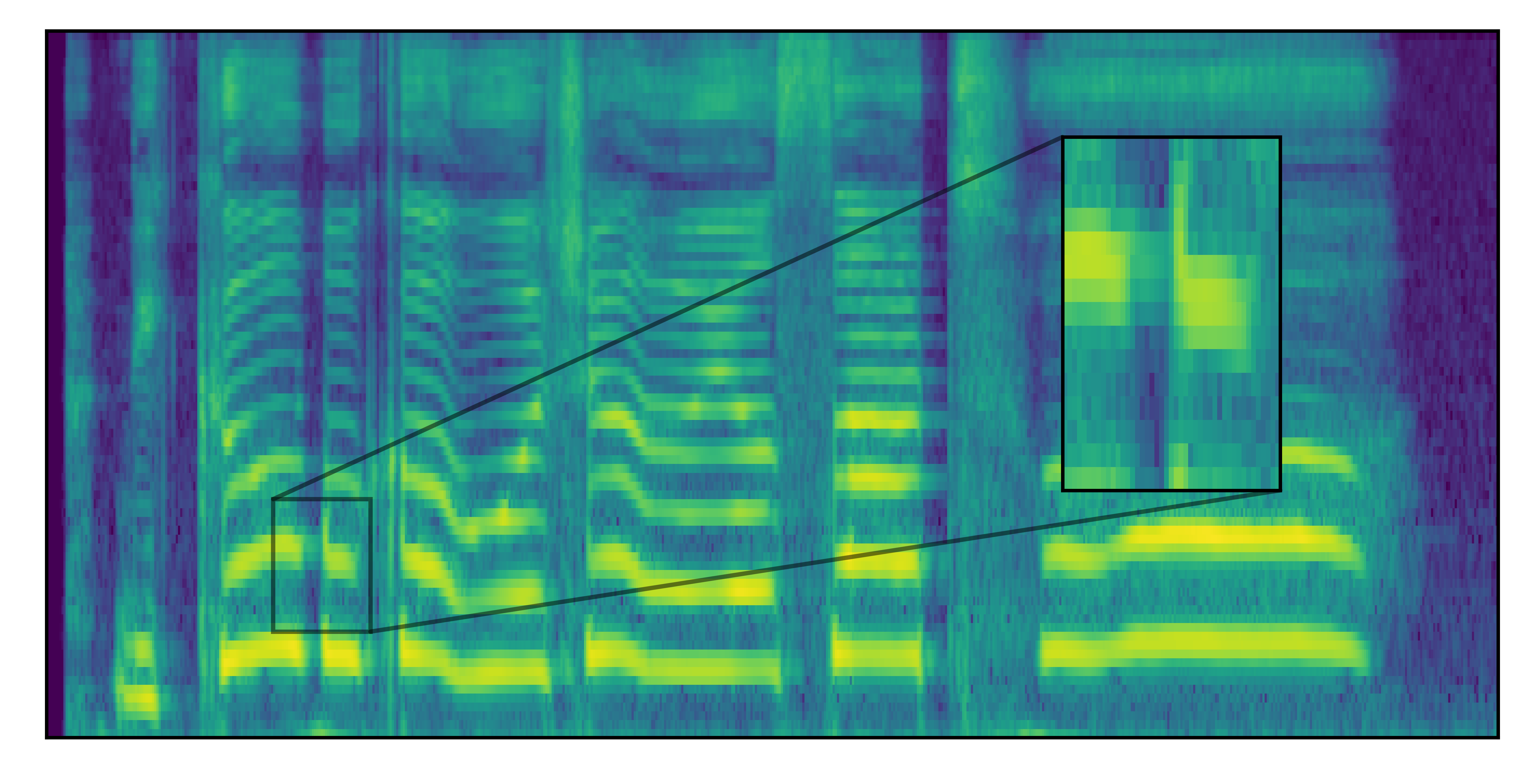}
        \caption{ \textit{GT}}
    	\label{fig:case_gt}
    	
    \end{subfigure}
    \begin{subfigure}{0.24\textwidth}
    	\centering
    	\includegraphics[width=\textwidth,trim={1.6cm 0.5cm 1.3cm 0.5cm},clip=true]{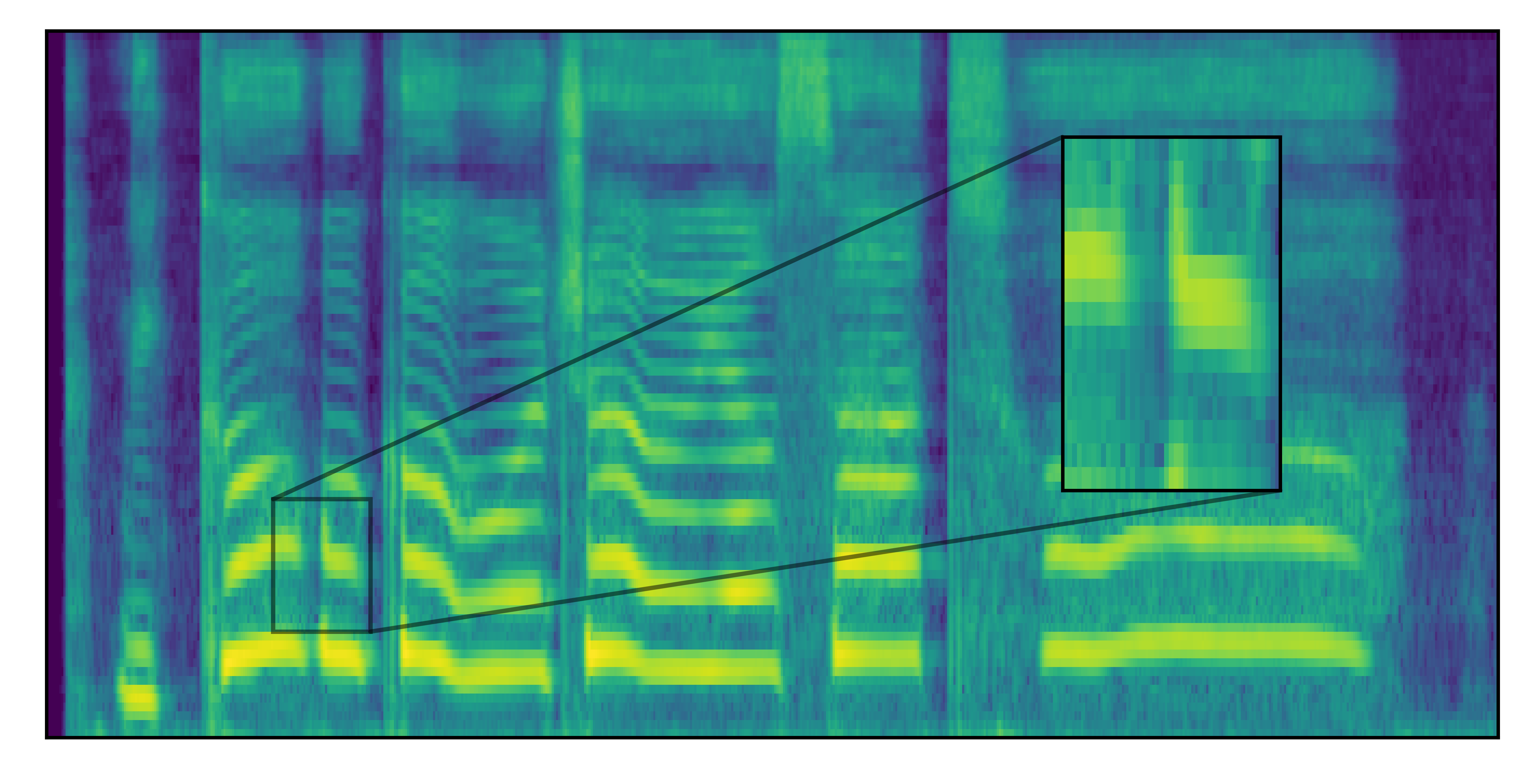}
        \caption{\textit{Diffsinger}}
    	\label{fig:case_diff}
    	
    \end{subfigure}
    \begin{subfigure}{0.24\textwidth}
    	\centering
    	\includegraphics[width=\textwidth,trim={1.6cm 0.5cm 1.3cm 0.5cm},clip=true]{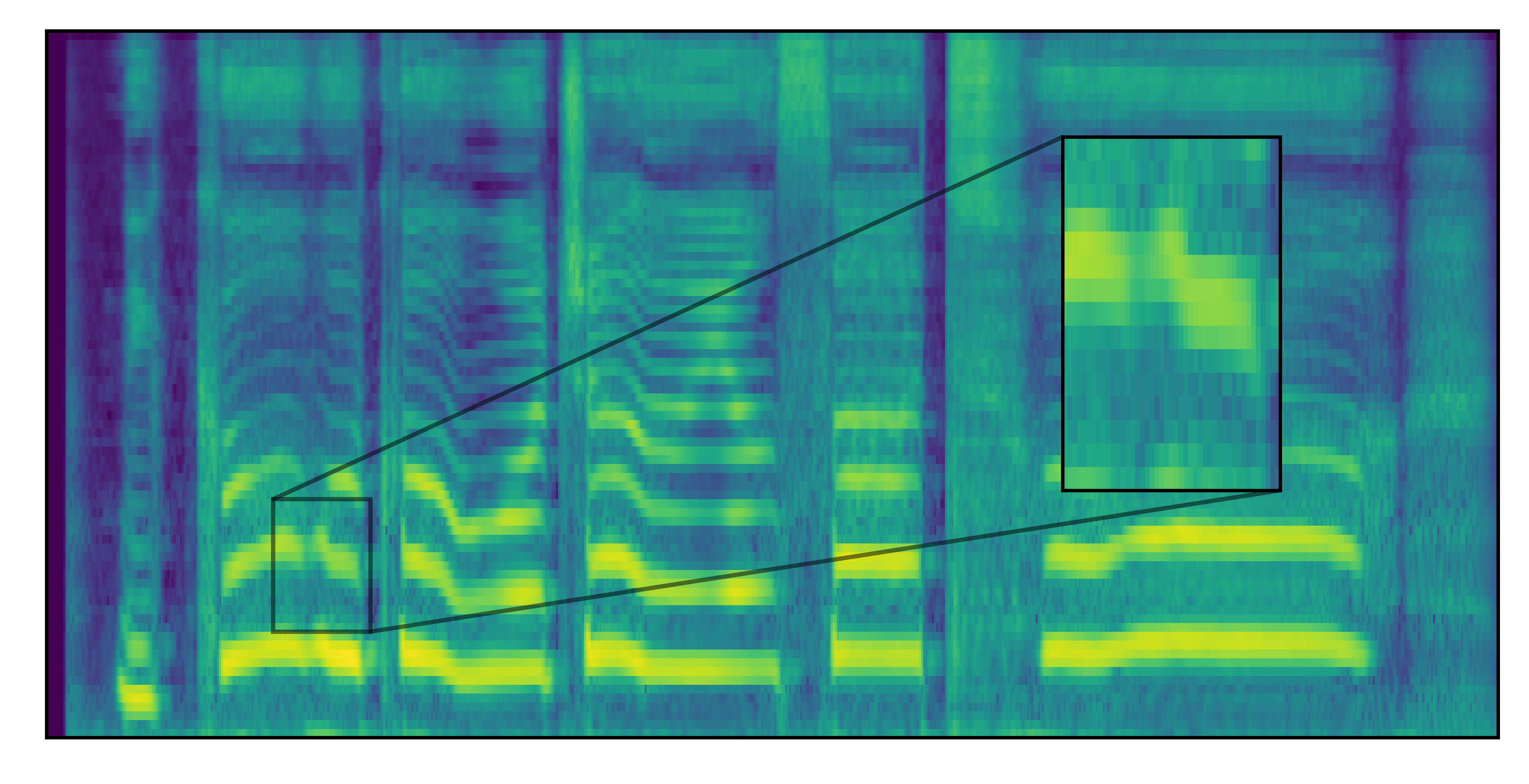}
        \caption{ \textit{GAN-singer}}
    	\label{fig:case_fsadv}
    	
    \end{subfigure}
    \begin{subfigure}{0.24\textwidth}
    	\centering
    	\includegraphics[width=\textwidth,trim={1.6cm 0.5cm 1.3cm 0.5cm},clip=true]{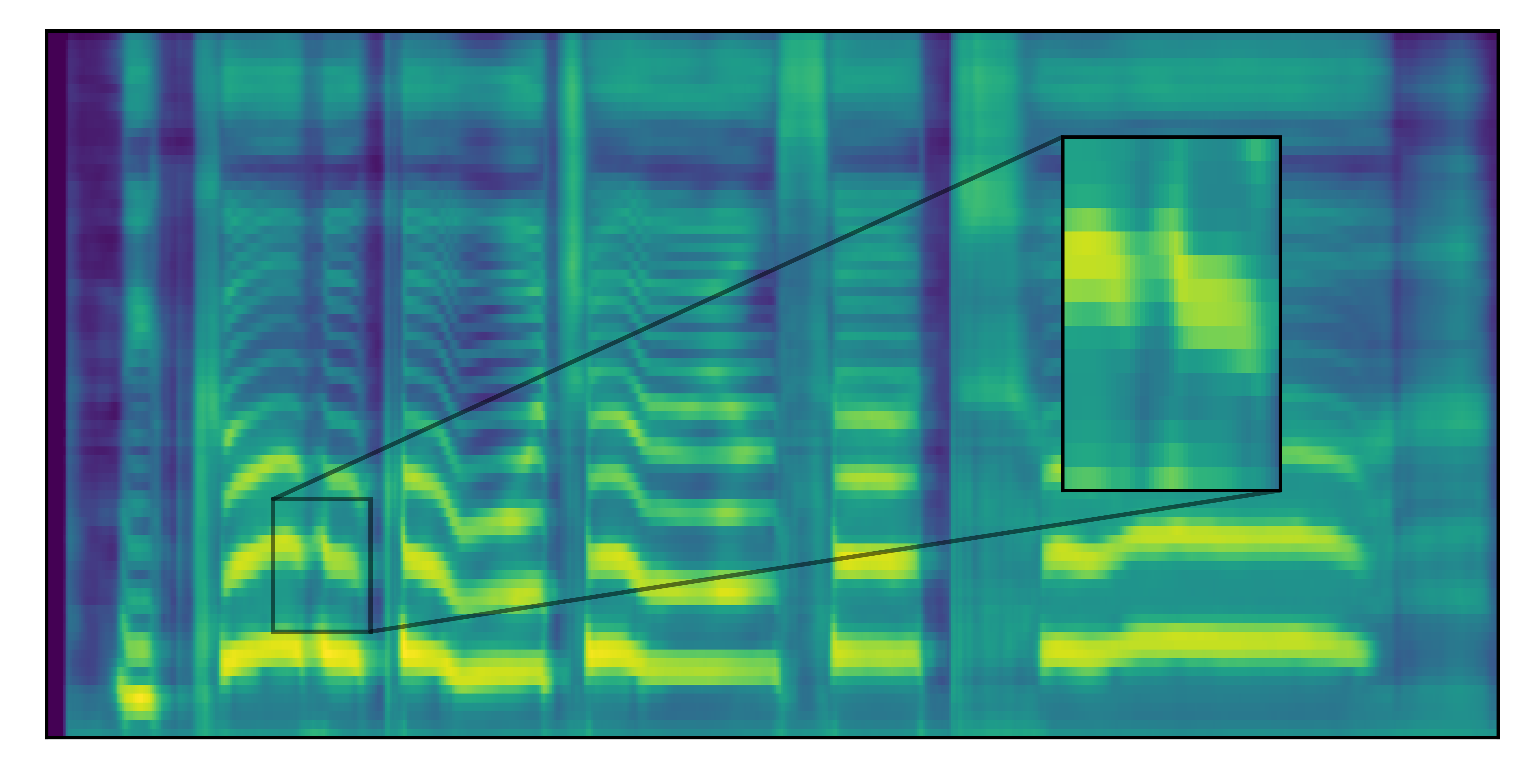}
        \caption{ \textit{FFT-Singer}}
    	\label{fig:case_fs2}
    	
    \end{subfigure}
    \caption{Visualizations of mel-spectrograms in four systems: GT, DiffSinger, GAN-Singer and FFT-Singer.}
    \label{fig:case_study}
\end{figure*}
\paragraph{Training and Inference}
\label{inference}
The training has two stages: 1) warmup stage: separately train the auxiliary decoder for 160k steps with the music score encoder, and then leverage the auxiliary decoder to train the boundary predictor for 30k steps to obtain $k$; 2) main stage: training DiffSinger as Algorithm~\ref{alg:training} describes for 160k steps until convergence. In the inference stage, for all SVS experiments, we uniformly use a pretrained Parallel WaveGAN (PWG)~\cite{yamamoto2020parallel}\footnote{We adjust PWG to take in F0 driven source excitation~\cite{wang2020using} as additional condition, similar to that in \cite{chen2020hifisinger}.} as vocoder to transform the generated mel-spectrograms into waveforms (audio samples).

\subsection{Main Results and Analysis} \label{sec:main_res}
\subsubsection{Audio Performance}
To evaluate the perceptual audio quality, we conduct the MOS (mean opinion score) evaluation on the test set. Eighteen qualified listeners are asked to make judgments about the synthesized song samples. We compare MOS of the song samples generated by DiffSinger with the following systems: 1) \textit{GT}, the ground truth singing audio; 2) \textit{GT (Mel + PWG)}, where we first convert the ground truth singing audio to the ground truth mel-spectrograms, and then convert these mel-spectrograms back to audio using PWG vocoder described in Section~\ref{inference}; 3) \textit{FFT-NPSS~\cite{blaauw2020sequence} (WORLD)}, the SVS system which generates WORLD vocoder features~\cite{morise2016world} through feed-forward Transformer (FFT) and uses WORLD vocoder to synthesize audio; 4) \textit{FFT-Singer (Mel + PWG)} the SVS system which generates mel-spectrograms through FFT network and uses PWG vocoder to synthesize audio; 5) \textit{GAN-Singer~\cite{wu2020adversarially} (Mel + PWG)}, the SVS system with adversarial training using multiple random window discriminators. 

The results are shown in Table~\ref{tab:main_exp}. The quality of \textit{GT (MEL + PWG)} (4.04 $\pm$ 0.11) is the upper limit of the acoustic model for SVS. \textit{DiffSinger} outperforms the baseline system with simple training loss (\textit{FFT-Singer}) by a large margin, and shows the superiority compared with the state-of-the-art GAN-based method (\textit{GAN-Singer}~\cite{wu2020adversarially}), which demonstrate the effectiveness of our method. 
\begin{table}[h]
    \small
	\centering
	\begin{tabular}{ l | c  }
		\toprule
		Method &  MOS  \\
		\midrule
		\textit{GT} & 4.30 $\pm$ 0.09  \\
	    \textit{GT (Mel + PWG)} & 4.04 $\pm$ 0.11  \\
		\midrule
		\textit{FFT-NPSS (WORLD)} & 1.75  $\pm$ 0.17  \\
		\textit{FFT-Singer (Mel + PWG)} & 3.67 $\pm$ 0.11 \\
		\textit{GAN-Singer (Mel + PWG)} & 3.74 $\pm$ 0.12  \\
		\midrule
		\textit{DiffSinger Naive (Mel + PWG)} & 3.71 $\pm$ 0.10 \\
		\textit{DiffSinger (Mel + PWG)} & \textbf{3.85} $\pm$ 0.11 \\
		\bottomrule
	\end{tabular}
	\caption{The MOS with 95\% confidence intervals of song samples. DiffSinger Naive means the naive version of DiffSinger without shallow diffusion mechanism.}
	\label{tab:main_exp}
\end{table}

As shown in Figure~\ref{fig:case_study}, we compare the ground truth, the generated mel-spectrograms from \textit{Diffsinger}, \textit{GAN-singer} and \textit{FFT-Singer} with the same music score. It can be seen that both Figure~\ref{fig:case_fsadv} and Figure~\ref{fig:case_diff} contain more delicate details between harmonics than Figure~\ref{fig:case_fs2} does. Moreover, the performance of \textit{Diffsinger} in the region of mid or low frequency is more competitive than that of \textit{GAN-singer} while maintaining similar quality of the high-frequency region.

In the meanwhile, the shallow diffusion mechanism accelerates inference of naive diffusion model by 45.1\% (RTF 0.191 vs. 0.348, RTF is the real-time factor, that is the seconds it takes to generate one second of audio). 
\subsubsection{Ablation Studies} \label{sec:ablation}
We conduct ablation studies to demonstrate the effectiveness of our proposed methods and some hyper-parameters studies to seek the best model configurations. We conduct CMOS evaluation for these experiments. The results of variations on \textit{DiffSinger} are listed in Table~\ref{tab:ablations}. It can be seen that: 1) removing the shallow diffusion mechanism results in quality drop (-0.500 CMOS), which is consistent with the MOS test results and verifies the effectiveness of our shallow diffusion mechanism (row 1 vs. row 2); 2) adopting other $k$ (row 1 vs. row 3) rather than the one predicted by our boundary predictor causes quality drop, which verifies that our boundary prediction network can predict a proper $k$ for shallow diffusion mechanism; and 3) the model with configurations $C=256$ and $L=20$ produces the best results (row 1 vs. row 4,5,6,7), indicating that our model capacity is sufficient.

\begin{table}[!h]
    \small
    \centering
    \begin{tabular}[width=\textwidth]{c|c|c|c|c|c}
    \toprule
    No. & $C$ & $L$ & w/ shallow & $k$ & CMOS \\
    \midrule
    1 & 256 & 20  & \checkmark  & 54 & 0.000 \\
    \midrule
    2 & 256 & 20  & $\times$    & -  & -0.500  \\  %
    \midrule
    3 & 256 & 20  & \checkmark  & 25 & -0.053  \\
    \midrule
    4 & 128 & 20  & \checkmark  & 54 & -0.071  \\
    5 & 512 & 20  & \checkmark  & 54 & -0.044  \\
    \midrule
    6 & 256 & 10  & \checkmark  & 54 & -0.293  \\
    7 & 256 & 30  & \checkmark  & 54 & -0.445  \\
    \bottomrule
    \end{tabular}
    \caption{Variations on the \textit{DiffSinger}. $T$ in all the experiments is set to 100. $C$ is channel size; $L$ is the number of layers in denoiser; w/ shallow means ``with shallow diffusion mechanism''; $k=54$ is our predicted intersection boundary.}
    \label{tab:ablations}
\end{table}

\subsection{Extensional Experiments on TTS} \label{sec:tts_exp}
\begin{table}[!h]
     
    \small

    \centering
    \begin{tabular}{ l | c }
    \toprule
    Method &  MOS \\
    \midrule
    \textit{GT} & 4.22 $\pm$ 0.07 \\
     \textit{GT (Mel + HiFi-GAN)} & 4.15 $\pm$ 0.07 \\
    \midrule
    \textit{Tacotron 2 (Mel + HiFi-GAN)} & 3.54  $\pm$ 0.05  \\
    \textit{BVAE-TTS (Mel + HiFi-GAN)} & 3.48 $\pm$ 0.06 \\
    \textit{FastSpeech 2 (Mel + HiFi-GAN)} & 3.68 $\pm$ 0.06 \\
    \textit{Glow-TTS (Mel + HiFi-GAN)} & 3.69 $\pm$ 0.07 \\
    \midrule
    \textit{DiffSpeech Naive (Mel + HiFi-GAN)} & 3.69 $\pm$ 0.05 \\
    \textit{DiffSpeech (Mel + HiFi-GAN)} & \textbf{3.92} $\pm$ 0.06 \\
    \bottomrule
    \end{tabular}
    \caption{The MOS of speech samples with 95\% confidence intervals.}
    \label{tab:exp_tts}
\end{table}
To verify the generalization of our methods on TTS task, we conduct the extensional experiments on LJSpeech dataset~\cite{ljspeech17}, which contains 13,100 English audio clips (total $\sim$24 hours) with corresponding transcripts. We follow the train-val-test dataset splits, the pre-processing of mel-spectrograms, and the grapheme-to-phoneme tool in FastSpeech 2. To build \textit{DiffSpeech}, we 1) add a pitch predictor and a duration predictor to \textit{DiffSinger} as those in FastSpeech 2; 2) adopt $k=70$ for shallow diffusion mechanism.

We use Amazon Mechanical Turk (ten testers) to make subjective evaluation and the results are shown in Table~\ref{tab:exp_tts}. All the systems adopt HiFi-GAN~\cite{kong2020hifi} as vocoder. \textit{DiffSpeech} outperforms FastSpeech 2 and Glow-TTS, which demonstrates the generalization. Besides, the last two rows in Table~\ref{tab:exp_tts} also show the effectiveness of shallow diffusion mechanism (with 29.2\% speedup, RTF 0.121 vs. 0.171).

\section{Related Work}
\subsection{Singing Voice Synthesis}
Initial works of singing voice synthesis generate the sounds using concatenated~\citep{macon1997concatenation,kenmochi2007vocaloid} or HMM-based parametric~\citep{saino2006hmm,oura2010recent} methods, which are kind of cumbersome and lack flexibility and harmony. Thanks to the rapid evolution of deep learning, several SVS systems based on deep neural networks have been proposed in the past few years. \citet{nishimura2016singing,blaauw2017neural,kim2018korean,nakamura2019singing,gu2020bytesing} utilize neural networks to map the contextual features to acoustic features. \citet{ren2020deepsinger} build the SVS system from scratch using singing data mined from music websites. \citet{blaauw2020sequence} propose a feed-forward Transformer SVS model for fast inference and avoiding exposure bias issues caused by autoregressive models. Besides, 
with the help of adversarial training, \citet{lee2019adversarially}~propose an end-to-end framework which directly generates linear-spectrograms. \citet{wu2020adversarially}~present a multi-singer SVS system with limited available recordings and improve the voice quality by adding multiple random window discriminators. \citet{chen2020hifisinger}~introduce multi-scale adversarial training to synthesize singing with a high sampling rate (48kHz). The voice naturalness and diversity of SVS system 
have been continuously improved in recent years.

\subsection{Denoising Diffusion Probabilistic Models}
A diffusion probabilistic model is a parameterized Markov chain trained by optimizing variational lower bound, which generates samples matching the data distribution in constant steps~\citep{Ho2020ddpm}. Diffusion model is first proposed by \citet{sohl2015deep}. \citet{Ho2020ddpm}~make progress of diffusion model to generate high-quality images using a certain parameterization and reveal an equivalence between diffusion model and denoising score matching~\citep{song2019generative,song2021scorebased}. Recently, \citet{kong2021diffwave}~and \citet{chen2021wavegrad}~apply the diffusion model to neural vocoders, which generate high-fidelity waveform conditioned on mel-spectrogram. \citet{chen2021wavegrad}~also propose a continuous noise schedule to reduce the inference iterations while maintaining synthesis quality. \citet{song2021denoising} extend diffusion model by providing a faster sampling mechanism, and a way to interpolate between samples meaningfully. Diffusion model is a fresh and developing technique, which has been applied in the fields of unconditional image generation, conditional spectrogram-to-waveform generation (neural vocoder). And in our work, we propose a diffusion model for the acoustic model which generates mel-spectrogram given music scores (or text). There is a concurrent work~\citep{jeong2021diff} at the submission time of our preprint which adopts a diffusion model as the acoustic model for TTS task.

\section{Conclusion}
\label{sec:conclu}
In this work, we proposed DiffSinger, an acoustic model for SVS based on diffusion probabilistic model. To improve the voice quality and speed up inference, we proposed a shallow diffusion mechanism. Specifically, we found that the diffusion trajectories of $M$ and $\widetilde{M}$ converge together when the diffusion step is big enough. Inspired by this, we started the reverse process at the intersection (step $k$) of two trajectories rather than at the very deep diffusion step $T$. Thus the burden of the reverse process could be distinctly alleviated, which improves the quality of synthesized audio and accelerates inference. The experiments conducted on PopCS demonstrate the superiority of DiffSinger compared with previous works, and the effectiveness of our novel shallow diffusion mechanism. The extensional experiments conducted on LJSpeech dataset prove the effectiveness of DiffSpeech on TTS task. The directly synthesis without vocoder will be future work.

\section*{Acknowledgments}
This work was supported in part by the National Key R\&D Program of China under Grant No.2020YFC0832505, No.62072397, Zhejiang Natural Science Foundation under Grant LR19F020006. Thanks participants of the listening test for the valuable evaluations.
\bibliography{aaai22}

\begin{thebibliography}{38}
\providecommand{\natexlab}[1]{#1}

\bibitem[{Blaauw and Bonada(2017)}]{blaauw2017neural}
Blaauw, M.; and Bonada, J. 2017.
\newblock A neural parametric singing synthesizer modeling timbre and
  expression from natural songs.
\newblock \emph{Applied Sciences}, 7(12): 1313.

\bibitem[{Blaauw and Bonada(2020)}]{blaauw2020sequence}
Blaauw, M.; and Bonada, J. 2020.
\newblock Sequence-to-sequence singing synthesis using the feed-forward
  transformer.
\newblock In \emph{ICASSP 2020-2020 IEEE International Conference on Acoustics,
  Speech and Signal Processing (ICASSP)}, 7229--7233. IEEE.

\bibitem[{Chen et~al.(2020)Chen, Tan, Luan, Qin, and Liu}]{chen2020hifisinger}
Chen, J.; Tan, X.; Luan, J.; Qin, T.; and Liu, T.-Y. 2020.
\newblock HiFiSinger: Towards High-Fidelity Neural Singing Voice Synthesis.
\newblock \emph{arXiv preprint arXiv:2009.01776}.

\bibitem[{Chen et~al.(2021)Chen, Zhang, Zen, Weiss, Norouzi, and
  Chan}]{chen2021wavegrad}
Chen, N.; Zhang, Y.; Zen, H.; Weiss, R.~J.; Norouzi, M.; and Chan, W. 2021.
\newblock WaveGrad: Estimating Gradients for Waveform Generation.
\newblock In \emph{International Conference on Learning Representations}.

\bibitem[{Gu et~al.(2020)Gu, Yin, Rao, Wan, Tang, Zhang, Chen, Wang, and
  Ma}]{gu2020bytesing}
Gu, Y.; Yin, X.; Rao, Y.; Wan, Y.; Tang, B.; Zhang, Y.; Chen, J.; Wang, Y.; and
  Ma, Z. 2020.
\newblock ByteSing: A Chinese Singing Voice Synthesis System Using Duration
  Allocated Encoder-Decoder Acoustic Models and WaveRNN Vocoders.
\newblock \emph{arXiv preprint arXiv:2004.11012}.

\bibitem[{He et~al.(2016)He, Zhang, Ren, and Sun}]{he2016deep}
He, K.; Zhang, X.; Ren, S.; and Sun, J. 2016.
\newblock Deep residual learning for image recognition.
\newblock In \emph{Proceedings of the IEEE conference on computer vision and
  pattern recognition}, 770--778.

\bibitem[{Ho, Jain, and Abbeel(2020)}]{Ho2020ddpm}
Ho, J.; Jain, A.; and Abbeel, P. 2020.
\newblock Denoising Diffusion Probabilistic Models.
\newblock In Larochelle, H.; Ranzato, M.; Hadsell, R.; Balcan, M.~F.; and Lin,
  H., eds., \emph{Advances in Neural Information Processing Systems},
  volume~33, 6832--6843. Curran Associates, Inc.

\bibitem[{Ito and Johnson(2017)}]{ljspeech17}
Ito, K.; and Johnson, L. 2017.
\newblock The LJ Speech Dataset.
\newblock \url{https://keithito.com/LJ-Speech-Dataset/}.
\newblock Accessed: 2019-10-12.

\bibitem[{Jeong et~al.(2021)Jeong, Kim, Cheon, Choi, and Kim}]{jeong2021diff}
Jeong, M.; Kim, H.; Cheon, S.~J.; Choi, B.~J.; and Kim, N.~S. 2021.
\newblock Diff-tts: A denoising diffusion model for text-to-speech.
\newblock \emph{arXiv preprint arXiv:2104.01409}.

\bibitem[{Kenmochi and Ohshita(2007)}]{kenmochi2007vocaloid}
Kenmochi, H.; and Ohshita, H. 2007.
\newblock Vocaloid-commercial singing synthesizer based on sample
  concatenation.
\newblock In \emph{Eighth Annual Conference of the International Speech
  Communication Association}.

\bibitem[{Kim et~al.(2018)Kim, Choi, Park, Kim, Kim, and Hahn}]{kim2018korean}
Kim, J.; Choi, H.; Park, J.; Kim, S.; Kim, J.; and Hahn, M. 2018.
\newblock Korean Singing Voice Synthesis System based on an LSTM Recurrent
  Neural Network.
\newblock In \emph{INTERSPEECH 2018}. ISCA.

\bibitem[{Kim et~al.(2020)Kim, Kim, Kong, and Yoon}]{kim2020glow}
Kim, J.; Kim, S.; Kong, J.; and Yoon, S. 2020.
\newblock Glow-TTS: A Generative Flow for Text-to-Speech via Monotonic
  Alignment Search.
\newblock \emph{Advances in Neural Information Processing Systems}, 33.

\bibitem[{Kong, Kim, and Bae(2020)}]{kong2020hifi}
Kong, J.; Kim, J.; and Bae, J. 2020.
\newblock HiFi-GAN: Generative Adversarial Networks for Efficient and High
  Fidelity Speech Synthesis.
\newblock \emph{Advances in Neural Information Processing Systems}, 33.

\bibitem[{Kong et~al.(2021)Kong, Ping, Huang, Zhao, and
  Catanzaro}]{kong2021diffwave}
Kong, Z.; Ping, W.; Huang, J.; Zhao, K.; and Catanzaro, B. 2021.
\newblock DiffWave: A Versatile Diffusion Model for Audio Synthesis.
\newblock In \emph{International Conference on Learning Representations}.

\bibitem[{Lee et~al.(2019)Lee, Choi, Jeon, Koo, and Lee}]{lee2019adversarially}
Lee, J.; Choi, H.-S.; Jeon, C.-B.; Koo, J.; and Lee, K. 2019.
\newblock Adversarially Trained End-to-End Korean Singing Voice Synthesis
  System.
\newblock \emph{Proc. Interspeech 2019}, 2588--2592.

\bibitem[{Macon et~al.(1997)Macon, Jensen-Link, George, Oliverio, and
  Clements}]{macon1997concatenation}
Macon, M.; Jensen-Link, L.; George, E.~B.; Oliverio, J.; and Clements, M. 1997.
\newblock Concatenation-based MIDI-to-singing voice synthesis.
\newblock In \emph{Audio Engineering Society Convention 103}. Audio Engineering
  Society.

\bibitem[{McAuliffe et~al.(2017)McAuliffe, Socolof, Mihuc, Wagner, and
  Sonderegger}]{mcauliffe2017montreal}
McAuliffe, M.; Socolof, M.; Mihuc, S.; Wagner, M.; and Sonderegger, M. 2017.
\newblock Montreal Forced Aligner: Trainable Text-Speech Alignment Using Kaldi.
\newblock In \emph{Interspeech}, 498--502.

\bibitem[{Morise, Yokomori, and Ozawa(2016)}]{morise2016world}
Morise, M.; Yokomori, F.; and Ozawa, K. 2016.
\newblock WORLD: a vocoder-based high-quality speech synthesis system for
  real-time applications.
\newblock \emph{IEICE TRANSACTIONS on Information and Systems}, 99(7):
  1877--1884.

\bibitem[{Nakamura et~al.(2019)Nakamura, Hashimoto, Oura, Nankaku, and
  Tokuda}]{nakamura2019singing}
Nakamura, K.; Hashimoto, K.; Oura, K.; Nankaku, Y.; and Tokuda, K. 2019.
\newblock Singing voice synthesis based on convolutional neural networks.
\newblock \emph{arXiv preprint arXiv:1904.06868}.

\bibitem[{Nichol and Dhariwal(2021)}]{nichol2021improved}
Nichol, A.~Q.; and Dhariwal, P. 2021.
\newblock Improved denoising diffusion probabilistic models.
\newblock In \emph{International Conference on Machine Learning}, 8162--8171.
  PMLR.

\bibitem[{Nishimura et~al.(2016)Nishimura, Hashimoto, Oura, Nankaku, and
  Tokuda}]{nishimura2016singing}
Nishimura, M.; Hashimoto, K.; Oura, K.; Nankaku, Y.; and Tokuda, K. 2016.
\newblock Singing Voice Synthesis Based on Deep Neural Networks.
\newblock In \emph{Interspeech}, 2478--2482.

\bibitem[{Oord et~al.(2016)Oord, Dieleman, Zen, Simonyan, Vinyals, Graves,
  Kalchbrenner, Senior, and Kavukcuoglu}]{vanwavenet}
Oord, A. v.~d.; Dieleman, S.; Zen, H.; Simonyan, K.; Vinyals, O.; Graves, A.;
  Kalchbrenner, N.; Senior, A.; and Kavukcuoglu, K. 2016.
\newblock WaveNet: A Generative Model for Raw Audio.
\newblock In \emph{9th ISCA Speech Synthesis Workshop}, 125--125.

\bibitem[{Oura et~al.(2010)Oura, Mase, Yamada, Muto, Nankaku, and
  Tokuda}]{oura2010recent}
Oura, K.; Mase, A.; Yamada, T.; Muto, S.; Nankaku, Y.; and Tokuda, K. 2010.
\newblock Recent development of the HMM-based singing voice synthesis
  system—Sinsy.
\newblock In \emph{Seventh ISCA Workshop on Speech Synthesis}.

\bibitem[{Ren et~al.(2021)Ren, Hu, Tan, Qin, Zhao, Zhao, and
  Liu}]{ren2021fastspeech}
Ren, Y.; Hu, C.; Tan, X.; Qin, T.; Zhao, S.; Zhao, Z.; and Liu, T.-Y. 2021.
\newblock FastSpeech 2: Fast and High-Quality End-to-End Text to Speech.
\newblock In \emph{International Conference on Learning Representations}.

\bibitem[{Ren et~al.(2020)Ren, Tan, Qin, Luan, Zhao, and
  Liu}]{ren2020deepsinger}
Ren, Y.; Tan, X.; Qin, T.; Luan, J.; Zhao, Z.; and Liu, T.-Y. 2020.
\newblock Deepsinger: Singing voice synthesis with data mined from the web.
\newblock In \emph{Proceedings of the 26th ACM SIGKDD International Conference
  on Knowledge Discovery \& Data Mining}, 1979--1989.

\bibitem[{Rethage, Pons, and Serra(2018)}]{rethage2018wavenet}
Rethage, D.; Pons, J.; and Serra, X. 2018.
\newblock A wavenet for speech denoising.
\newblock In \emph{2018 IEEE International Conference on Acoustics, Speech and
  Signal Processing (ICASSP)}, 5069--5073. IEEE.

\bibitem[{Saino et~al.(2006)Saino, Zen, Nankaku, Lee, and
  Tokuda}]{saino2006hmm}
Saino, K.; Zen, H.; Nankaku, Y.; Lee, A.; and Tokuda, K. 2006.
\newblock An HMM-based singing voice synthesis system.
\newblock In \emph{Ninth International Conference on Spoken Language
  Processing}.

\bibitem[{Shen et~al.(2018)Shen, Pang, Weiss, Schuster, Jaitly, Yang, Chen,
  Zhang, Wang, Skerrv-Ryan et~al.}]{shen2018natural}
Shen, J.; Pang, R.; Weiss, R.~J.; Schuster, M.; Jaitly, N.; Yang, Z.; Chen, Z.;
  Zhang, Y.; Wang, Y.; Skerrv-Ryan, R.; et~al. 2018.
\newblock Natural tts synthesis by conditioning wavenet on mel spectrogram
  predictions.
\newblock In \emph{ICASSP 2018}, 4779--4783. IEEE.

\bibitem[{Sohl-Dickstein et~al.(2015)Sohl-Dickstein, Weiss, Maheswaranathan,
  and Ganguli}]{sohl2015deep}
Sohl-Dickstein, J.; Weiss, E.; Maheswaranathan, N.; and Ganguli, S. 2015.
\newblock Deep Unsupervised Learning using Nonequilibrium Thermodynamics.
\newblock In \emph{International Conference on Machine Learning}, 2256--2265.

\bibitem[{Song, Meng, and Ermon(2021)}]{song2021denoising}
Song, J.; Meng, C.; and Ermon, S. 2021.
\newblock Denoising Diffusion Implicit Models.
\newblock In \emph{International Conference on Learning Representations}.

\bibitem[{Song and Ermon(2019)}]{song2019generative}
Song, Y.; and Ermon, S. 2019.
\newblock Generative Modeling by Estimating Gradients of the Data Distribution.
\newblock In \emph{Proceedings of the 33rd Annual Conference on Neural
  Information Processing Systems}.

\bibitem[{Song et~al.(2021)Song, Sohl-Dickstein, Kingma, Kumar, Ermon, and
  Poole}]{song2021scorebased}
Song, Y.; Sohl-Dickstein, J.; Kingma, D.~P.; Kumar, A.; Ermon, S.; and Poole,
  B. 2021.
\newblock Score-Based Generative Modeling through Stochastic Differential
  Equations.
\newblock In \emph{International Conference on Learning Representations}.

\bibitem[{Vaswani et~al.(2017)Vaswani, Shazeer, Parmar, Uszkoreit, Jones,
  Gomez, Kaiser, and Polosukhin}]{vaswani2017attention}
Vaswani, A.; Shazeer, N.; Parmar, N.; Uszkoreit, J.; Jones, L.; Gomez, A.~N.;
  Kaiser, {\L}.; and Polosukhin, I. 2017.
\newblock Attention is all you need.
\newblock In \emph{Advances in Neural Information Processing Systems},
  5998--6008.

\bibitem[{Vincent(2011)}]{score2011vincent}
Vincent, P. 2011.
\newblock A Connection Between Score Matching and Denoising Autoencoders.
\newblock In \emph{Neural Computation}.

\bibitem[{Wang and Yamagishi(2020)}]{wang2020using}
Wang, X.; and Yamagishi, J. 2020.
\newblock Using Cyclic Noise as the Source Signal for Neural
  Source-Filter-Based Speech Waveform Model.
\newblock \emph{Proc. Interspeech 2020}, 1992--1996.

\bibitem[{Wu and Luan(2020)}]{wu2020adversarially}
Wu, J.; and Luan, J. 2020.
\newblock Adversarially Trained Multi-Singer Sequence-to-Sequence Singing
  Synthesizer.
\newblock \emph{Proc. Interspeech 2020}, 1296--1300.

\bibitem[{Yamamoto, Song, and Kim(2020)}]{yamamoto2020parallel}
Yamamoto, R.; Song, E.; and Kim, J.-M. 2020.
\newblock Parallel WaveGAN: A fast waveform generation model based on
  generative adversarial networks with multi-resolution spectrogram.
\newblock In \emph{ICASSP 2020-2020 IEEE International Conference on Acoustics,
  Speech and Signal Processing (ICASSP)}, 6199--6203. IEEE.

\bibitem[{Zhang et~al.(2020)Zhang, Yu, Lu, Weng, Zhang, Wu, Xie, Li, and
  Yu}]{zhang2020durian}
Zhang, L.; Yu, C.; Lu, H.; Weng, C.; Zhang, C.; Wu, Y.; Xie, X.; Li, Z.; and
  Yu, D. 2020.
\newblock DurIAN-SC: Duration Informed Attention Network Based Singing Voice
  Conversion System.
\newblock \emph{Proc. Interspeech 2020}, 1231--1235.

\end{thebibliography}
\appendix
\section{Theoretical Proof of Intersection}
\label{sup:proof}
Given a data sample $M_0$ and its corresponding $\widetilde{M}_0$, the conditional distributions of $M_t$ and $\widetilde{M}_t$ are:
\begin{align*}
&q(M_t| M_0)=\mathcal{N}(M_t; \sqrt{\bar\alpha_t}M_0, (1-\bar\alpha_t)\mathbf{I}) \\
&q(\widetilde{M}_t| \widetilde{M}_0)=\mathcal{N}(\widetilde{M}_t;\sqrt{\bar\alpha_t}\widetilde{M}_0, (1-\bar\alpha_t)\mathbf{I})
\end{align*}
respectively. The KL-divergence between two Gaussian distributions is:
\begin{equation*}
\small
\begin{split}
D_{\mathrm{KL}}\left(\mathcal{N}_{0} \| \mathcal{N}_{1}\right)=\frac{1}{2}[\operatorname{tr}\left(\Sigma_{1}^{-1} \Sigma_{0}\right)+\left(\mu_{1}-\mu_{0}\right)^{\top} \Sigma_{1}^{-1}\left(\mu_{1}-\mu_{0}\right) \\
-k+\ln (\frac{\operatorname{det} \Sigma_{1}}{\operatorname{det} \Sigma_{0}})]
\end{split}
\end{equation*}
where $k$ is the dimension; $\mu_{0}, \mu_{1}$ are means; $\Sigma_{0}, \Sigma_{1}$ are covariance matrices. Thus, in our case:
\begin{equation*}
D_{KL}(\mathcal{N}(M_t) || \mathcal{N}(\widetilde{M}_t)) = \frac{\bar\alpha_t}{2(1-\bar\alpha_t)} \|\widetilde{M}_0-M_0 \| _2^2
\end{equation*}
Since $\frac{\bar\alpha_t}{2(1-\bar\alpha_t)}$ decreases towards 0 rapidly as $t$ increases, this KL-divergence also decreases towards 0 rapidly as $t$ increases. This guarantees the intersection of trajectories of the diffusion process. 

Moreover, since the auxiliary decoder has been optimized by simple reconstruction loss (L1/L2 mentioned in the main paper) on the training set, $\|\widetilde{M}_0-M_0 \| _2^2$ is optimized towards minimum, which facilitates this intersection. In addition, $\widetilde{M}_k$ does not need to be exactly the same as $M_k$, but just needs to come from vicinity of the mode of $q(M_k|M_0)$ (according to the theories of score matching and Langevin dynamics). 

\section{An Easier Trick for Boundary Prediction}
Intuitively, we can just adopt the smallest $t$ as $k$ when $t$ satisfies:
\begin{align*}
    &\Eb{M\in \mathcal{Y'}}{D_{\mathrm{KL}}\left(\mathcal{N}(M_t)  \| \mathcal{N}(\widetilde{M}_t))\right)} \\ 
    &=  \Eb{M\in \mathcal{Y'}}{\frac{\bar\alpha_t}{2(1-\bar\alpha_t)} \|\widetilde{M}_0-M_0 \| _2^2}\\ &\leq \Eb{M\in \mathcal{Y'}}{D_{KL}(\mathcal{N}(M_T) || \mathcal{N}(\bzero, \bI))},
\end{align*}
which means that this start point at step $k$ is at least not worse than the original prior distribution $\mathcal{N}(\bzero, \bI)$. $\mathcal{Y'}$ mean the mel-spectrograms in the validation set. In addition, when using this trick to determine $k$ in DiffSpeech, it is more rational to generate $\widetilde{M}_t$ ($\in \mathcal{Y'}$) conditioned on the ground-truth F0-contour \& duration rather than the ones predicted.

\section{Details of Model Structure and Supplementary Configurations}
\label{sup:model_details}
\subsection{Details of model structure}
\begin{figure}[!h]
    \centering
    \begin{subfigure}{0.24\textwidth}
    	\centering
    	\includegraphics[width=\textwidth,clip=true]{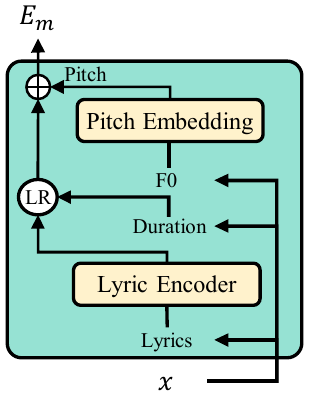}
    	\caption{Encoder.}
    	\label{supfig:encoder}
    \end{subfigure}
    \begin{subfigure}{0.22\textwidth}
    	\centering
    	\includegraphics[width=\textwidth,clip=true]{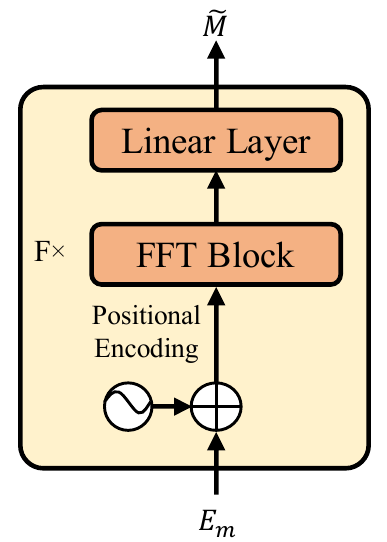}
    	\caption{Auxiliary Decoder.}
    	\label{supfig:aux_decoder}
    \end{subfigure}
    \caption{The detailed model structure of Encoder and Auxiliary Decoder. $x$ is the music score. $E_m$ is the music condition sequence. $\widetilde{M}$ means the blurry mel-spectrogram generated by the auxiliary decoder trained with L1 loss. }
    \label{supfig:encoder_auxdecoder}
\end{figure}

\begin{figure}[!h]
	\centering
	\small
	\includegraphics[width=0.45\textwidth,]{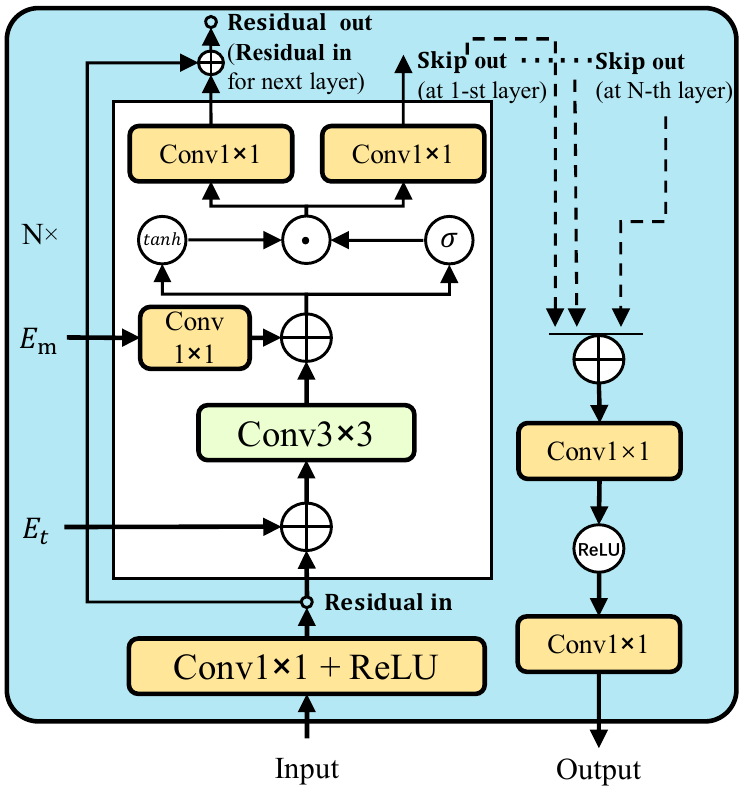}
	\caption{The detailed model structure of Denoiser. $E_t$ is step embedding and $E_m$ is music condition sequence. $N$ is the number of residual layers. The model structure is derived from non-causal WaveNet, but simplified by replacing dilation layer to naive convolution layer.}
	\label{supfig:denoiser}
\end{figure}

The detailed model structure of encoder, auxiliary decoder and denoiser are shown in Figure~\ref{supfig:encoder}, Figure~\ref{supfig:aux_decoder} and Figure~\ref{supfig:denoiser} respectively. 

\subsection{Supplementary configurations}
In each FFT block: the number of FFT layers (F in Figure~\ref{supfig:aux_decoder}) is set to 4; the hidden size of self-attention layer is 256; the number of attention heads is 2; the kernel sizes of 1D-convolution in the 2-layer convolutional layers are set to 9 and 1.

\section{Model Size}
The model footprints of main systems for comparison in our paper are shown in Table~\ref{suptab:model_size}. It can be seen that DiffSinger has the similar learnable parameters as other state-of-the-art models. 
\begin{table}[!h]
\begin{center}
    \begin{tabular}{ l | c } 
        \toprule
        \textbf{Model} &  \textbf{Param(M)} \cr
        \midrule 
        \multicolumn{1}{l}{\textit{SVS Models}}         \\
        \midrule 
        DiffSinger & 26.744 \cr
        \midrule
        FFT-Singer & 24.254 \cr
        \midrule
        \multirow{2}*{GAN-Singer} 
         ~ & 24.254 (Generator)  \cr
         ~ & 0.963 (Discriminator)  \cr
        \midrule
        \multicolumn{1}{l}{\textit{TTS Models}}         \\
        \midrule
        DiffSinger & 27.722 \cr
        \midrule
        Tacotron 2 & 28.193 \cr
        \midrule
        BVAE-TTS & 15.991 \cr
        \midrule
        FastSpeech 2 & 24.179 \cr
        \midrule
        Glow-TTS & 28.589 \cr
        \midrule
    \end{tabular}
\end{center}
\caption{The model footprints. Param means the learnable parameters.}
\label{suptab:model_size}
\end{table}

\section{Details of Training and Inference}
We train DiffSinger on 1 NVIDIA V100 GPU with 48 batch size. We adopt the Adam optimizer with learning rate $lr = 10^{-3}$. During training, the warmup stage costs about 16 hours and the main stage costs about 12 hours; During inference, the RTF of acoustic model for SVS and TTS are 0.191 and 0.121 respectively.

\end{document}